\documentclass[twocolumn,prl]{revtex4}
\usepackage{amsfonts}

\usepackage{graphicx}
\usepackage{epsfig}
\usepackage{amsmath, amssymb}
\usepackage{verbatim}
\usepackage{bm}
\usepackage{multirow}

\newcommand{\rchi}{{\mathpalette\irchi\relax}}
\newcommand{\irchi}[2]{\raisebox{\depth}{$#1\chi$}}
\newcommand{\rgamma}{{\mathpalette\irgamma\relax}}
\newcommand{\irgamma}[2]{\raisebox{\depth}{$#1\gamma$}}

\begin{document}

\title{Classification of topological defects in Abelian topological states}
\author{Maissam Barkeshli}
\author{Chao-Ming Jian}
\author{Xiao-Liang Qi}
\affiliation{Department of Physics, Stanford University, Stanford, CA 94305 }

\begin{abstract}
In this paper we propose the most general classification of point-like and line-like extrinsic topological defects in
$(2+1)$-dimensional Abelian topological states. We first map generic extrinsic defects to boundary defects, and then provide
a classification of the latter. Based on this classification, the most generic point defects can be understood as domain walls
between topologically distinct boundary regions. We show that topologically distinct boundaries can themselves be classified by certain maximal
subgroups of mutually bosonic quasiparticles, called Lagrangian subgroups. We study the topological properties of the point defects,
including their quantum dimension, localized zero modes, and projective braiding statistics.
\end{abstract}
\maketitle

A fundamental discovery in condensed matter physics has been the
understanding of topologically ordered states of matter\cite{wen04,nayak2008}. Topologically ordered states
possess quasiparticle excitations with fractional statistics, topology-dependent
ground state degeneracies, and long-range entanglement, all of which are robust even without symmetry.
The most common topological orders seen experimentally are the fractional quantum Hall (FQH) states, while
there is increasing evidence that they may be observed in frustrated magnets\cite{balents2010}

\begin{figure}
\centerline{
\includegraphics[width=2.8in]{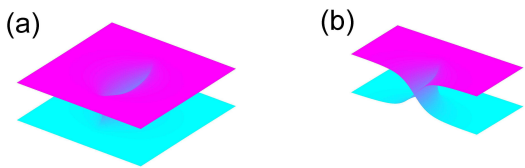}
}
\caption{\label{fig:genon} (a) Schematic picture of bilayer system with a pair of genons.\cite{barkeshli2012a,barkeshli2013genon} (b) is the same figure as (a) with part of the system removed, to see the branch-cut line clearly.
}
\end{figure}

Recently, a new aspect of topologically ordered states, called twist defects or extrinsic defects,
has attracted increasing research interest. \cite{barkeshli2010,bombin2010,
barkeshli2012a,barkeshli2013genon,barkeshli2013,you2012,you2012b,clarke2013,lindner2012,cheng2012,vaezi2013,
brown2013,kitaev2012}
An extrinsic defect is a point-like or line-like defect either in a topological state, or on the interface between two
topologically distinct states, which leads to topological properties that are absent without the defect.
A simple example is a ``genon" \cite{barkeshli2010,barkeshli2012a,barkeshli2013genon}:
Consider a branch-cut line in a bilayer topological state, across which the two
layers are exchanged (Fig. \ref{fig:genon}). A genon is defined as an end point of the branch cut.
It was observed that the bilayer system with genons is topologically equivalent to a single layer system on
a high genus surface, yielding a topological degeneracy that grows exponentially with the number of genons, and
a notion of (projective) braiding statistics that can be studied systematically \cite{barkeshli2013genon}.
Even when the topological state in each layer is Abelian, the genons have non-Abelian statistics.
This has led to a recent experimental proposal for realizing a wide class of topological qubits
in conventional bilayer FQH states\cite{barkeshli2013}, and an understanding of how to realize
universal topological quantum computation in non-universal, non-Abelian states \cite{barkeshli2013genon}.
Extrinsic defects with the same type of non-Abelian statistics as genons have also been proposed in other physical systems, such
as lattice defects in certain exactly solvable $Z_N$ rotor models,\cite{you2012,you2012b} and FQH %or fractional quantum spin Hall (FQSH)
states in proximity with superconductivity (SC) and ferromagnetism (FM). \cite{fu2009d,lindner2012,clarke2013,cheng2012}

\begin{figure}
\centerline{
\includegraphics[width=3in]{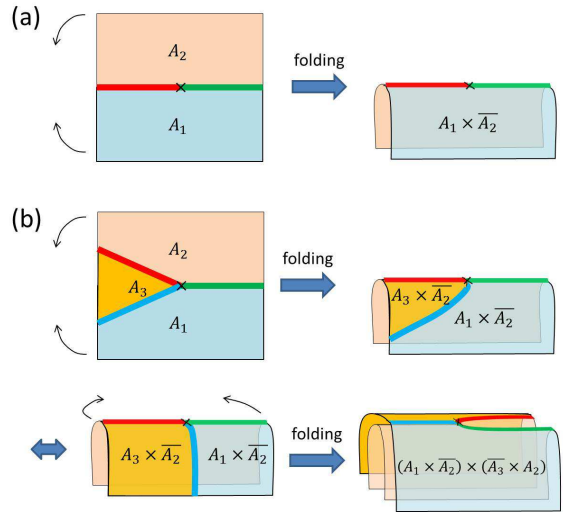}
}
\caption{\label{pointFig} (a) A domain wall between two different kinds of gapped edges separating
topological phases $A_1$ and $A_2$. By folding $A_2$ over, this can be mapped to a domain wall on the boundary separating
$A_1 \times \bar{A}_2$ and the vacuum. (b) A junction where multiple gapped edges meet is
also a possible type of point defect. On an infinite plane, by applying the folding trick multiple times, this can also be
mapped to a domain wall on the boundary separating a topological phase and the vacuum.
}
\end{figure}

In this paper, we report a general theory of extrinsic defects in Abelian topological states, extending
the theory of the extrinsic defects reviewed above to the most general possible form. 
For two-dimensional topological states, there are two general forms of extrinsic defects: \it Line defects\rm, which separate two different or identical topological
states, and \it point defects\rm, which may exist in a single topological state, such as twist defects \cite{barkeshli2013genon},  or
at junctions between different line defects (Fig. \ref{pointFig}). We demonstrate that all extrinsic defects can be mapped to {\it boundary defects}, {\it i.e.},
boundary lines of topological states with point defects separating different boundary regions.
Based on the understanding of boundary defects, we develop a classification of gapped line defects between Abelian topological states, extending previous results
\cite{bravyi1998,kapustin2010,kapustin2011,beigi2011,kitaev2012,wang2012,levin2013}.
We prove that gapped line defects are classified by
``Lagrangian subgroups," which consist of subgroups of topological quasi-particles that have trivial self and mutual
statistics, and that are condensed on the boundary.

The nontrivial point defects on the boundary are then classified by domain walls between topologically distinct line defects. 
We obtain the quantum dimension of general point defects, demonstrating that they are
generally non-Abelian and can be understood in terms of the fractional statistics of bulk quasiparticle excitations.
We show that the point defects localize a set of topologically protected zero modes,
which can be understood as a localized, robust non-zero density of states at zero energy for a certain subgroup of the
topological quasiparticles. Finally, we will briefly discuss the sense in which these non-Abelian boundary defects
can be braided.

\bf Abelian topological states and line defects --\rm Abelian topological states in 2+1 dimensions are generically described by
Abelian Chern-Simons (CS) theories\cite{wen04,wen1995}:
$\mathcal{L}_{CS} = \frac{1}{4\pi} K_{IJ} \epsilon^{\mu\nu\lambda} a^I_\mu \partial_\nu a^J_\lambda $,
where $a^I$ for $I = 1, \cdots, N$ are compact $U(1)$ gauge fields, and $K$ is a non-singular,
integer symmetric matrix.
A topological quasiparticle carries point charges $l_I\in\mathbb{Z}$ of $a^I$. The statistics of a quasiparticle
labeled by the integer vector $l$ is given by $\theta_l = \pi l^T K^{-1} l $, and the mutual statistics of two quasiparticles
$l$, $l'$ is $\theta_{ll'} = 2\pi l^T K^{-1} l'$. A quasiparticle with $l_I=K_{IJ}v_J,~v_J\in\mathbb{Z}$ is considered as a local
``electron" in the theory, which may be bosonic or fermionic depending on $K$. Therefore the topologically nontrivial
quasi-particles are labeled by integer vectors $l$ mod $Kv$, % which forms a lattice on an $N$-dimensional torus,
with the number of topologically distinct quasiparticles given by $|\text{Det }K|$.

Different $K$-matrices can specify equivalent topological states if they have the same
quasiparticle content. For example, $K' = W^T K W$, for $W$
an integer matrix with $|\text{Det } W| = 1$, describes the same
topological order. Another example is
\begin{eqnarray}
K'=K\oplus P\label{expandK}
\end{eqnarray}
with $P$ an integer matrix with $|{\rm Det~}P |=1$.
%$\tau_x=\left(\begin{array}{cc}0&1\\1&0\end{array}\right)$ the Pauli matrix. Since ${\rm Det~}\tau_x=1$,
Adding $P$ does not introduce any new topological quasiparticles, so that $K$ and $K'$ describe the same topological order.

A general line defect in a topological state is a one-dimensional boundary between two topological states,
$A_1$ and $A_2$ (see Fig. \ref{pointFig} (a)). Some line defects, such as the edge of chiral topological states,
are robustly gapless\cite{wen04,haldane1995,levin2013}. In this work we will explore gapped line
defects.

In order to understand the properties of general boundaries, it is helpful to apply a folding process\cite{beigi2011,kapustin2011}
(see Fig. \ref{pointFig} (a)).
By folding the upper half plane using a parity transformation relative to the line defect, $A_2$ is
mapped to its parity conjugate $\bar{A}_2$, so that the line defect
becomes a boundary between the topological state $A_1 \times \bar{A}_2$ and a topologically trivial gapped state.
Therefore, to study gapped line defects, it suffices to consider all possible gapped boundaries between general
topological phases and the trivial state.

\bf Classification of gapped boundaries-- \rm The key feature of a gapped boundary of a topological
phase is that some subgroup of the topological quasiparticles are \it condensed \rm on the boundary,
and can be created/annihilated on the boundary by \it local \rm operators \cite{levin2013}. Physically this
describes superselection sectors for how topological quasiparticles can be reflected/transmitted at line defects \cite{beigi2011,kitaev2012}.
We first consider the genon case \cite{barkeshli2012a,barkeshli2013genon} as an example.

\begin{figure}
\centerline{
\includegraphics[width=3in]{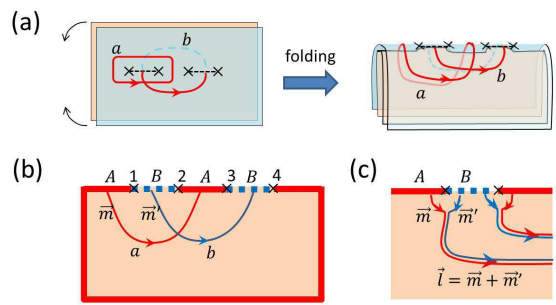}
}
\caption{\label{fig:genonloops}
(a) The non-contractible loops in the bilayer system with $4$ genons. Loop $a$ is in the upper (blue) layer and loop $b$ runs from upper layer to the lower (orange) layer across the branch-cut lines. After folding, the genons become domain wall between different gapped boundaries, and the non-contractible loops become Wilson lines that terminate on the boundaries. (b) The general Wilson lines in a system with boundary defects and point defects. Lines $a$ ($b$) defines the unitary operator $W_{\bf m}(a)$ ($W_{\bf m'}(b)$) which correspond to adiabatic motion of bosonic quasiparticle ${\bf m}({\bf m'})$ along the paths $a(b)$, respectively.
}
\end{figure}

Consider a simple bilayer topological state, the $(mm0)$ Halperin state\cite{halperin1983}, with the
$K$ matrix $K=m\mathbb{I}_{2\times 2}$, which describes two independent $1/m$-Laughlin FQH states. Here $\mathbb{I}_{2\times 2}$ is the $2$-dimensional identity matrix.
%Here $\mathbb{I}_{2\times 2}$ denotes the $2\times 2$ identity matrix.
Folding the state along a line (see Fig. \ref{fig:genonloops} (a)) %including a genon
we obtain a 4-layer system with the $K$ matrix
%\begin{eqnarray}
$K=\left(\begin{array}{cc}m\mathbb{I}_{2\times 2}&0\\0&-m\mathbb{I}_{2\times 2}\end{array}\right)$. %\label{Kmm0}
%\end{eqnarray}
The boundary of such a state can be gapped by introducing either %, as there are an equal number of counterpropagating chiral edge states that can backscatter into each other. We consider two distinct types of boundary conditions, which correspond to
interlayer or intralayer backscattering. The genon is defined as the domain wall between these two types of boundaries.

These two boundary conditions can be distinguished by the behavior of quasiparticles at the boundary.
Across the boundary gapped by intralayer backscattering, quasiparticles move between layers $1,\bar{1}$
and $2,\bar{2}$, so that quasiparticles of the type $l=(q_1,q_2,-q_1,-q_2)^T$ can be annihilated or created
at the boundary. Such quasiparticles have bosonic self-statistics and mutual-statistics, and thus can be
considered to be ``condensed" on the boundary. Similarly, across the boundary defined by interlayer backscattering, a different
set of quasiparticles with  $l=(q_1,q_2,-q_2,-q_1)^T$ are condensed.

We see that different gapped boundaries condense different subgroups of quasiparticles.
In general, it has been proven that every gapped boundary must condense a subgroup of quasiparticles
$M$, called a ``Lagrangian subgroup," which has the following properties \cite{levin2013}\footnote{For bosonic states,
a similar subgroup was referred to as a Lagrangian subgroup in \cite{kapustin2011}, but it was not proven that gapped boundaries
must condense such a subgroup.}:
\begin{enumerate}
\item $e^{i \theta_{mm'}} = 1$ for all $m$, $m'$ $\in M$;
\item For all $l\notin M$, $e^{i \theta_{lm}} \neq 1$ for at least one $m \in M$.
\end{enumerate}
For bosonic states (when all diagonals of $K$ are even), we also have $e^{i \theta_m} = 1$ for all $m \in M$. This set of
quasiparticles form an Abelian group with the group multiplication defined by particle fusion. The first condition
defines the bosonic mutual statistics and bosonic or fermionic self-statistics, allowing $m\in M$ to be condensed on the boundary.
The second condition guarantees that the boundary is completely gapped, since all other quasiparticles $l\notin M$
have nontrivial mutual statistics with particles in $M$, and thus are confined when the quasiparticles in $M$ are condensed.

In the following we will strengthen this result by proving that \it every \rm Lagrangian subgroup $M$ corresponds to a gapped boundary
where $M$ is condensed. 

To explicitly write down the boundary condition corresponding to a Lagrangian subgroup,
we introduce the edge theory of the CS theory $\mathcal{L}_{CS}$ defined above, which is given by
the chiral Luttinger liquid theory\cite{wen04,wen1995} 
%\begin{eqnarray}
%\label{edgeL}
$\mathcal{L}_{edge} = \frac{1}{4\pi} K_{IJ}\partial_x \phi_I \partial_t \phi_J - V_{IJ} \partial_x \phi_I \partial_x \phi_J $.
%\end{eqnarray}
$V_{IJ}$ is a real symmetric positive definite matrix, and $\phi_I$ are real compact scalar fields: $\phi_I \sim \phi_I + 2\pi$.
If $K$ has an equal number of positive and negative eigenvalues, then there are an equal number of left- and right- moving modes,
which is a necessary but not sufficient condition for the edge to be gapped.

The electron annihilation operators $\Psi_I$ and quasiparticle annihilation operators $\rchi_l$ on the boundary are given by
%\begin{eqnarray}
$\Psi_I = e^{i K_{IJ} \phi_J}, \;\; \rchi_l = e^{i l_I \phi_I}$,
%\end{eqnarray}
where $l$ is an integer vector describing the quasiparticles. Naively, the condensation of a quasiparticle $m\in M$
can be described by adding a term $\frac g2(\rchi_{m}+\rchi_{m}^\dagger)=g\cos \left(m_{I}\phi_I\right)$ to $\mathcal{L}_{edge}$.
However, such a term has two problems. First, it is not a local term, written in terms of local ``electron'' operators $\Psi_I$.
Secondly, the condition $m^TK^{-1}m=0$ must be satisfied in order for the phase $m_{I}\phi_I$ to obtain a classical value.
With this condition, it is possible to perform a change of basis $\phi \rightarrow W \phi$ so that the theory is mapped to a standard non-chiral
Luttinger liquid, with  $\cos \left(m_{I}\phi_I\right)$ mapped to a conventional backscattering term.
%({\bf Explain more?})
The first problem can be solved by multiplying an integer coefficient $c_i$, such that
$c_iK^{-1}m_i\equiv \Lambda_i\in \mathbb{Z}$, and thus
$\cos \left(c_im_{iI}\phi^I\right)=\cos \left(\Lambda_i^TK\phi\right)$ is a local electron tunneling operator.

The second can be solved if we can find a set of generators $\{m_i\}$ of $M$ satisfying $m_i^TKm_j=0,~\forall i,~j$. Then,
the term $g\sum_i\cos \left(c_im_{iI}\phi_I \right)$ can be added to the Lagrangian and will condense the particles
in $M$: $\langle e^{i m^T \phi} \rangle \neq 0$ if $m \in M$. It is known that if one can find $N$ such null vectors
$m_i$ for a $2N\times 2N$ $K$ matrix, the edge can be completely gapped\cite{haldane1995}.

However, it is not always possible to find such a null vector basis $\{m_i\}$ which fully generates $M$.
For example, consider $K = \left(\begin{matrix} 0 & 4 \\ 4 & 0 \end{matrix} \right)$,
which describes $Z_4$ topological order. This system has a Lagrangian subgroup generated by
$m_1^T = (2,0)$, $m_2^T = (0,2)$. It is not possible to find a single null vector which generates
this Lagrangian subgroup. Consequently, it is not clear what $\cos$ term on the boundary
leads to the condensation of this Lagrangian subgroup.

This problem can be resolved by introducing a topologically equivalent $K$ matrix with
higher dimension, as is shown in Eq. (\ref{expandK}). In the edge theory, adding additional
trivial blocks $P$ such as $P=\tau_x$ or $\tau_z$, where $\tau_i$ are $2\times 2$ Pauli matrices,
corresponds to adding purely one-dimensional edge channels to the
boundary, such as Heisenberg spin $1/2$ chains. Thus we find:

\noindent{{\bf Lemma}: For each Lagrangian subgroup $M$ of the topological state described
by $K$, there exists a $K'$  which is topologically equivalent to $K$ and has
${\rm rank}(K')=2N'$, such that the same Lagrangian subgroup $M$ of $K'$ can be generated by $N'$ null vectors $m_i',~i=1,2,...,N'$.}

The proof of this conclusion will be presented in the appendix. As a simple example that illustrates the main
idea of the proof, consider the previous example, with
$K =  \left(\begin{matrix} 0 & 4 \\ 4 & 0 \end{matrix} \right)$, and
$m_1^T = (2,0)$, $m_2^T = (0,2)$. We define
$K' = \left(\begin{matrix} K & 0 \\ 0 & \tau_x \end{matrix} \right)$, and
$m_1'^T = (2,0,0,1)$, $m_2'^T = (0,2,-1,0)$. Here, $K'$ is topologically equivalent
to $K$, and $m_i'^T K'^{-1} m_j' = 0$. Physically, this result implies that one can always condense
the particles in a Lagrangian subgroup on the edge,
%by pinning appropriate cosine terms to their classical minima in the Luttinger liquid theory,
as long as using additional trivial edge states is allowed.

\it We conclude that every Lagrangian subgroup $M$ corresponds to a gapped boundary where $M$
is condensed, providing a classification, in the absence of any symmetries, of topologically distinct
gapped boundaries. \rm

%\noindent{
\bf Classification and characterization of point defects-- \rm In general, a point defect
is a junction where multiple different line defects meet. Under the folding process, which may be applied
multiple times, the point defects can always be mapped to domain walls between two gapped edges (Fig. \ref{pointFig}).
Therefore it is sufficient to study the point defect at the domain wall between two gapped boundaries. Based on the
above classification of gapped boundaries, %in terms of Lagrangian subgroups,
the domain walls are thus classified by a pair of Lagrangian subgroups $(M, M')$, corresponding to the gapped boundaries on either
side of the domain wall.

Consider a point defect labelled by $(M, M')$.
In the genon example reviewed above, the simplest topological property of the point defect is its nontrivial
quantum dimension. This can be understood from the fact that the bilayer system (on the sphere) with $2n$ genons has
genus $n-1$, which leads to a topological ground state degeneracy that grows exponentially in $n$.
For Abelian states, the topological degeneracy can be obtained from the algebra of the Wilson loop operators,
which measure the topological charge through non-contractible loops.  For example, a sphere with
$4$ genons is equivalent to a torus, which has two non-contractible loops $a,b$ (see Fig. \ref{fig:genonloops} (a)).
When each layer is a $1/m$ Laughlin state, the Wilson loop operators $W(a)$ and $W(b)$ are defined by creating a pair of charge $1/m,-1/m$ particles
and taking one of them around the loops $a$ and $b$, and then annihilating them. $W(a)$ and $W(b)$ satisfy the
commutation relation $W(a)W(b)=W(b)W(a)e^{i2\pi/m}$, and each leave the system in its ground state subspace,
requiring the ground state degeneracy to be an integer multiple of $m$.\cite{barkeshli2013genon}.

These Wilson loop operators can be generalized to the generic point defects. By folding the bilayer system
with genons along a line containing the genons, the Wilson loops become Wilson lines of the particles
$(1/m,0,-1/m,0)$ or $(1/m,0,0,-1/m)$, which terminate at the boundary since the corresponding particles
are condensed at the boundary. For general defects, between two gapped boundaries $A,B$ with
Lagrangian subgroups $M$ and $M'$, respectively, the Wilson lines can be defined (Fig. \ref{fig:genonloops}
(b)) by creating a boson $m\in M$ at the $A$ boundary using a local operator, moving it along a path $a$ connecting two $A$
regions, and finally annihilating it using a local operator. We denote such an operator as $W_m(a)$ and similarly $W_{m'}(b)$ for
moving particle $m'\in M'$ between two $B$ regions. The commutation relation between $W_m(a)$ and
$W_{m'}(b)$ is determined by the mutual statistics of particles $m$ and $m'$, which is nontrivial when
$M$ and $M'$ are different Lagrangian subgroups: \begin{eqnarray}
\label{wilsonalgebra}
W_m(a)W_{m'}(b)=W_{m'}(b)W_{m}(a)e^{2\pi im^TK^{-1}m'}.
\end{eqnarray}
Since these operators leave the system in the ground state subspace, the ground states must form a representation
of this algebra.

The degeneracy $D$ required by one pair of non-contractible intersecting lines $a,b$ is
the dimension of the minimal representation of the algebra (\ref{wilsonalgebra}), which can be obtained by acting one set of operators, such as $\{W_m(a)\}$,
on the eigenstates of the other set $\{W_{m'}(b)\}$. 
On a boundary with $2n$ defects between $n$ pairs of alternating  $A$ and $B$ regions, there will be $n-1$ pairs of non-commuting line operators
satisfying the same algebra as above, leading to a degeneracy $D^{n-1}$. Therefore each point defect has a
quantum dimension of $d=\sqrt{D}$.

If fermions exist microscopically in the system, there may be an additional $\sqrt{2}$ factor in the quantum dimension, originating from
the Majorana zero modes of purely one-dimensional physics \cite{kitaev2001}, which is independent of the above analysis.

\bf Localized zero modes \rm --A key feature of the point defects is that they localize a non-zero density of states at zero energy
for a certain subgroup of quasiparticles. Such zero modes have been studied for specific types of defects\cite{barkeshli2013genon,lindner2012,clarke2013}, and here we show that they exist in general point defects. Consider a point defect at $x=0$ between two boundary regions $A$ at $x<0$ and $B$ at $x>0$, which are labeled by Lagrangian subgroups $M$ and $M'$. For quasiparticles $m\in M,~m'\in M'$, the boson creation operators $\rchi_m(-\epsilon)=e^{im_I\phi_I(-\epsilon)}$ and $\rchi_{m'}(\epsilon)=e^{im'_I\phi_I(\epsilon)}$ for $\epsilon>0$ create condensed quasi-particles in $A$ and $B$ regions correspondingly. Therefore the operator $\rchi_m(-\epsilon)\rchi_{m'}(\epsilon)$ preserves the ground state manifold. Taking the limit $\epsilon\rightarrow 0$ we obtain a local operator at the point defect:
\begin{eqnarray}
\rgamma_{l}\equiv\lim_{\epsilon\rightarrow 0^+}\rchi_m(-\epsilon)\rchi_{m'}(\epsilon)\label{zeromode}
\end{eqnarray}
with $l=m+m'$. By construction, bilinear combinations of $\rgamma_{l}$ on different defects preserve the ground state manifold, which means $\rgamma_l$ is a zero mode operator. The zero mode creation process can be understood as the emission of a quasi-particle $l$ which has fractional statistics, as is illustrated in Fig. \ref{fig:genonloops} (c).
The zero modes $\rgamma_l$ are generalizations of the parafermion zero modes \cite{fradkin1980,zamolodchikov1985,fendley2012}
studied in previous works %In the special cases studied in
\cite{barkeshli2013genon,barkeshli2013,lindner2012,clarke2013,cheng2012}.

\bf Effective braiding of point-defects \rm-- In the systems with genons or more
generic twist defects, the non-Abelian braid statistics of defects can be defined projectively\cite{barkeshli2013genon},
which enables topologically protected transformations on the degenerate ground states.
The more general boundary defects studied here cannot always be braided in real space,
since the defects are confined to the boundary line defects. However,
it has been proposed\cite{alicea2010,lindner2012,clarke2013,cheng2012,barkeshli2013genon,bonderson2013}
that effective braiding operations can be achieved by controlling the quasiparticle tunneling between point defects.
The general defects we studied can be coupled by tunneling of the zero mode quasiparticles $\rgamma_l$ in Eq. (\ref{zeromode}).
The tunneling Hamiltonian has the form $H_{ab} = \sum_{l \in L} t_l \rgamma_l^\dagger(x_a) \rgamma_l(x_b) + H.c.$, where $x_a$ and $x_b$ are the positions
of two equivalent defects, $t_l$ are tunneling amplitudes, and $\rgamma_l(x_a)$ are the zero mode operators at the defects.
If $H_{ab}$ has a unique ground state in the Hilbert space defined by the two defects $a,b$, one can consider an adiabatic deformation of
Hamiltonian $H(\lambda)=H_{ab}(1-\lambda)+H_{bc}\lambda$. When $\lambda$ adiabatically changes from $0$ to $1$, the defect at
$c$ is transferred to $a$. Combining such motions of defects, braiding between two defects can be realized. However, in order for
$H_{ab}$ to have a unique ground state in a Hilbert space defined by $a,b$, we need the set $L$ appearing in $H_{ab}$ to consist only of quasiparticles
of the form $l_i=m_i+m_i'$, with the highly non-trivial requirement that $m_i^TK^{-1}m_j'=m_j^TK^{-1}m_i'$, where $\{m_i\}$ ($\{m_i'\}$) is a set of generators of the Lagrangian
subgroup $M$ ($M'$).  Otherwise we have not been able to define a notion of braiding. A general analysis of such braiding statistics for generic defects will be presented in later work \cite{barkeshli2013defect}.

\noindent{\bf Acknowledgement.} We thank Alexei Kitaev for helpful discussions. After finishing this work, we
learned that some of the results on the classification of line defects are independently found by Michael Levin
and included in an updated version of Ref. \cite{levin2013}. We acknowledge support from the Simons
foundation (MB) and Packard Foundation (XLQ).

%\bibliography{TI}

\newpage

\appendix

\section{Proof of Lemma}

Here we prove the following lemma, which was presented in the main text:

\noindent{{\bf Lemma}: For each Lagrangian subgroup $M$ of the topological state described
by $K$, there exists a $K'$  which is topologically equivalent to $K$ and has
${\rm rank}(K')=2N'$, such that the same Lagrangian subgroup $M$ of $K'$ can be generated by $N'$ null vectors $m_i',~i=1,2,...,N'$.}

To prove this, we will focus on two cases independently. In the first case, $K$ has only even entries along the diagonals,
which describes topological phases where the microscopic degrees of freedom only consist of bosons. In the second case,
$K$ can have odd entries along the diagonals, which is appropriate when the microscopic degrees of freedom have at least
one species of fermions.

We note that our Lemma and proof builds on results presented in an early version of Ref. \cite{levin2013}.
As our paper was about to appear, we learned that the Lemma and proof presented here was independently 
also found by M. Levin and included in an updated version of Ref. \cite{levin2013}.

\subsection{Proof for $K$ even}

Let us first consider the case where $K$ is an even matrix, meaning that its diagonal entries are all even. Note that
$K$ is also an integer symmetric non-singular matrix with vanishing signature. Consider the lattice
\begin{align}
\Gamma = \{m + K\Lambda: m \in M, \Lambda \in \mathbb{Z}^{2N}\}.
\end{align}
$\Gamma$ is a $2N$-dimensional integer lattice, and can be written as $\Gamma = U \mathbb{Z}^{2N}$, where
$U$ is a $2N$-dimensional integer matrix. Define:
\begin{align}
P = U^T K^{-1} U.
\end{align}
$P$ is an even integer symmetric matrix with unit determinant and non-vanishing signature\cite{levin2013}. The fact that it is
an even integer matrix follows because the columns of $U$ generate the Lagrangian subgroup $M$, and these
all have bosonic mutual and self-statistics by definition. The fact that it is symmetric and has vanishing signature follows
from the fact that $K$ is symmetric and has vanishing signature. Finally, $P$ has unit determinant for the following
reason. Consider any integer vector $\Lambda \in \mathbb{Z}^{2N}$, and any non-integer $2N$-component vector, $x$.
By definition of the Lagrangian subgroup, $\Lambda^T P x$ must be non-integer, which implies that $P x$ must be non-integer.
This then implies that if $P x$ is integer for any $2N$-component vector $x$, then $x$ must be integer, which in turn implies that
$P^{-1}$ is integer. $P$ and $P^{-1}$ can both be integer if and only if $P$ has unit determinant.

Since $P$ is an even symmetric integer matrix with vanishing signature and unit determinant,
it follows from a mathematical theorem \cite{milnor1973}
that it is always possible to find a $SL(2N;\mathbb{Z})$ transformation $W$ such that
\begin{align}
W^T P W = \left(\begin{matrix} 0 & \mathbb{I} \\ \mathbb{I} & 0 \end{matrix} \right),
\end{align}
where $I$ is an $N \times N$ identity matrix.
Thus, we consider a transformed theory:
\begin{align}
\tilde{U} &= W^T U W,
\nonumber \\
\tilde{K} &= W^T K W,
\nonumber \\
\tilde{P} &= W^T P W =  \left(\begin{matrix} 0 & \mathbb{I} \\ \mathbb{I} & 0 \end{matrix} \right)
\end{align}
Since $W \in SL(2N;\mathbb{Z})$, $\tilde{K}$ and $K$ describe topologically equivalent theories.
Clearly, the columns of $\tilde{U}$ generate the Lagrangian subgroup $M$.
Let $\tilde{u}_{i}$ denote the $i$th column of $\tilde{U}$. Let us extend $\tilde{K}$
to a $4N \times 4N$ matrix $K'$, which is composed of $\tilde{K}$ and $N$ copies of
$\tau_x = \left(\begin{matrix} 0 & 1 \\ 1 & 0 \end{matrix} \right)$
along the block diagonal entries:
\begin{align}
K' = \left(\begin{matrix} \tilde{K} &  & & \\
 & \tau_x &  & \\
 &  & \tau_x & \\
&  & & \ddots\end{matrix} \right),
\end{align}
where the rest of the entries are zero. Again, $K'$ describes the same
topological order as $K$. Now we define
\begin{align}
\vec{m}_1'^T &= (\tilde{u}_1^T, 0, 1, 0, 0, \cdots,0,0)
\nonumber \\
\vec{m}_2'^T &= (\tilde{u}_{N+1}^T, -1, 0, 0,0, \cdots,0,0)
\nonumber \\
\vec{m}_3'^T &= (\tilde{u}_2^T,0,0,0,1,\cdots,0,0)
\nonumber \\
\vec{m}_4'^T &= (\tilde{u}_{N+2}^T,0,0,-1,0,\cdots,0,0)
\nonumber \\
\vdots
\nonumber \\
\vec{m}_{2N-1}'^T &= (\tilde{u}_{N}^T,0,0,\cdots,0,1)
\nonumber \\
\vec{m}_{2N}'^T &= (\tilde{u}_{2N}^T,0,0,\cdots,-1,0)
\end{align}
Since the additional components added in $K'$ are all trivial degrees of freedom, the $2N$ vectors
$\{\vec{m}_i'\}$ still generate the same Lagrangian subgroup $M$. It is easy to see that
\begin{align}
m_i'^T K'^{-1} m'_j = 0.
\end{align}
This proves the lemma for $K$ even. In practice, in most cases of interest it is easy to find $N$ columns
of $\tilde{U}$ that generate $M$ and that satisfy $\tilde{u}^T_i K^{-1} \tilde{u}_j = 0$, so the
above extension to a $4N$ dimensional $K$-matrix will not be necessary.

\subsection{Proof for $K$ odd}

Let us now consider the case where $K$ is odd (\it ie \rm it has at least one odd element along
the diagonal). As before, we define the matrix $U$, and $P = U^T K^{-1} U$. Now, $P$ is an
integer symmetric, non-singular matrix with unit determinant, non-vanishing signature, and at
least one odd element along the diagonal. Under these conditions, it is always possible to find
$W \in SL(2N;\mathbb{Z})$ such that \cite{milnor1973}
\begin{align}
W^T P W = \left(\begin{matrix} \mathbb{I} & 0 \\ 0 & -\mathbb{I} \end{matrix} \right),
\end{align}
where $I$ is an $N \times N$ identity matrix.
Thus, we consider a transformed theory:
\begin{align}
\tilde{U} &= W^T U W,
\nonumber \\
\tilde{K} &= W^T K W,
\nonumber \\
\tilde{P} &= W^T P W =  \left(\begin{matrix} \mathbb{I} & 0 \\ 0 & -\mathbb{I} \end{matrix} \right).
\end{align}
Again, the original Lagrangian subgroup is generated by the columns of $\tilde{U}$.

Let $\tilde{u}_{i}$ denote the $i$th column of $\tilde{U}$, and
$\tilde{u}_{i\pm} = \tilde{u}_i \pm \tilde{u}_{N+i}$, for $i = 1, \cdots, N$.

Now, as in the case where $K$ is even, let us extend the $K$-matrix to a
$4N \times 4N$ matrix $K'$, which is composed of $K$ and now with $N$ copies of
$\tau_z = \left(\begin{matrix} 1 & 0 \\ 0 & -1 \end{matrix} \right)$
along the block diagonal entries:
\begin{align}
K' = \left(\begin{matrix} \tilde{K} &  & & \\
 & \tau_z &  & \\
 &  & \tau_z & \\
&  & & \ddots\end{matrix} \right),
\end{align}
where the rest of the entries are zero. In the absence of any symmetries,
$K'$ describes the same topological order as $K$.
Now we define
\begin{align}
\vec{m}_1'^T &= (\tilde{u}_{1+}^T, 1, 1, 0, 0, \cdots,0,0)
\nonumber \\
\vec{m}_2'^T &= (\tilde{u}_{1-}^T, -1, 1, 0,0, \cdots,0,0)
\nonumber \\
\vec{m}_3'^T &= (\tilde{u}_{2+}^T,0,0,1,1,\cdots,0,0)
\nonumber \\
\vec{m}_4'^T &= (\tilde{u}_{2-}^T,0,0,-1,1,\cdots,0,0)
\nonumber \\
\vdots
\nonumber \\
\vec{m}_{2N-1}'^T &= (\tilde{u}_{N+}^T,0,0,\cdots,1,1)
\nonumber \\
\vec{m}_{2N}'^T &= (\tilde{u}_{N-}^T,0,0,\cdots,-1,1)
\end{align}
Since the additional components added to $K'$ are all trivial degrees of freedom, the $2N$ vectors
$\{\vec{m}_i'\}$ still generate the same Lagrangian subgroup $M$. It is easy to see that
\begin{align}
m_i'^T K'^{-1} m'_j = 0.
\end{align}
This proves the lemma for $K$ odd.

\end{document}